\documentclass[twocolumn,english,superscriptaddress,pra]{revtex4-1}

\usepackage{amssymb,amsmath,latexsym,amscd,amsfonts, bbm, bm}  
\usepackage{graphicx}  

\usepackage{umoline}
\usepackage[usenames,svgnames]{xcolor}

\usepackage{hyperref}
\hypersetup{
     colorlinks=true,           
     linkcolor=Salmon,          
     citecolor=blue,            
     filecolor=blue,            
     urlcolor=cyan,             
 }

\newtheorem{theorem}{Theorem}[section]

\newtheorem{lemma}[theorem]{Lemma}

\newcommand{\Eq}[1]{Eq.~(\ref{#1})}
\newcommand{\Fig}[1]{Fig.~\ref{#1}}

\newcommand{\Lem}[1]{Lemma~\ref{#1}}

\newcommand{\Ref}[1]{Ref.~\cite{#1}}

\newcommand{\Thm}[1]{Theorem~\ref{#1}}
\newcommand{\App}[1]{Appendix~\ref{#1}}

\newcommand{\qedsymb}{\hfill{\rule{2mm}{2mm}}}

\newcommand{\ignore}[1]{}

\newcommand{\norm}[1]{{\| #1 \|}}  
  
\newcommand{\ket}[1]{{ |{#1} \rangle }}  
\newcommand{\bra}[1]{{ \langle {#1} | }}
\newcommand{\av}[1]{{ \langle {#1} \rangle }}
\newcommand{\braket}[2]{{ \langle {#1} | {#2} \rangle}}

\newcommand{\EqDef}{\stackrel{\mathrm{def}}{=}}
\newcommand{\Id}{\mathbbm{1}}

\newcommand{\etal}{\textit{et al}}

\DeclareMathOperator*{\Tr}{Tr}

\newcommand{\cc}[1]{~\cite{#1}}
\newcommand{\Bx}{\bm{x}}
\newcommand{\Bz}{\bm{z}}

\newcommand{\mcT}{\mathcal{T}}
\newcommand{\mcF}{\mathcal{F}}
\newcommand{\mcE}{\mathcal{E}}
\newcommand{\mcV}{\mathcal{V}}
\newcommand{\mcG}{\mathcal{G}}
\newcommand{\mcN}{\mathcal{N}}

\begin{document}

\title{Tensor Networks contraction and the Belief Propagation
algorithm}

\author{R.~Alkabetz}
\author{I.~Arad}
\affiliation{Department of Physics, Technion, 3200003 Haifa, Israel}

\date{\today}

\begin{abstract}
  Belief Propagation is a well-studied message-passing algorithm
  that runs over graphical models and can be used for approximate
  inference and approximation of local marginals. The resulting
  approximations are equivalent to the Bethe-Peierls
  approximation of statistical mechanics.  Here we show how this
  algorithm can be adapted to the world of PEPS tensor networks and used
  as an approximate contraction scheme. We further
  show that the resultant approximation is equivalent to the ``mean
  field'' approximation that is used in the Simple-Update algorithm,
  thereby showing that the latter is a essentially the Bethe-Peierls
  approximation. This shows that one of the simplest approximate
  contraction algorithms for tensor networks is equivalent to one of
  the simplest schemes for approximating marginals in graphical
  models in general, and paves the way for using improvements of BP
  as tensor networks algorithms.
\end{abstract}

\maketitle

\section{Introduction}

There is a natural connection between classical probabilistic
systems of many random variables and quantum many-body systems. In
both cases the description of a generic state of a system requires
an exponential number of parameters in the size of the system, or
the number of physical units that compose it. For example, a general
probability distribution over $n$ bits requires the specification of
$2^n-1$ non-negative numbers, while a classical description of a
quantum state over $n$ qubits requires $2^n-1$ complex numbers.
However, in both cases, states that are relevant to us are often
subject to many local constraints, which, in turn may lead to a
succinct description of the system. A good example is tensor
networks (TNs)\cc{ref:Orus2014-TN}, where the $2^n$ coefficients of
a quantum state are given by the contraction of a set of local
tensors. As a probabilistic analog, consider the Gibbs distribution
of a classical spin system on a lattice $H=\sum_{\av{a,b}}
h_{ab}(x_a, x_b)$. Also here, the multivariate probability
distribution $P(x_1, \ldots, x_n) = \frac{1}{Z} e^{-\beta H(x_1,
\ldots, x_n)} = \frac{1}{Z}\prod_{\av{a,b}}e^{-\beta h_{ab}(x_a,
x_b)}$ can be compactly described by specifying the local
interactions $h_{ab}(x_a, x_b)$. This is an example of a
\emph{graphical model}\cc{ref:Wainwright2008-GM, ref:Koller2009-PGM,
ref:Mezard2009-Info}, in which the global probability distribution
is given by a product of local factors. Tensor networks and
graphical models are therefore two frameworks that provide a compact
description of the state of the system, which in principle can be
used to simulate it.

Both frameworks also face similar challenges. In both cases, 
calculating the expectation value of a local observable can be an
NP-hard problem\cc{ref:Cooper1990-BPcomp,
ref:Schuch2007-PEPS-complexity}, as it (at least naively) involves
summation over an exponential number of terms. In addition, when the
underlying graph that describes the model is a tree, this can be
done efficiently using dynamical programming (via the sum-product
algorithm\cc{ref:Wainwright2008-GM} for graphical models or directly
by the results of \Ref{ref:Markov2008-TreeWidth} for
tensor-networks), but when there are loops, the problem becomes hard
and one usually resorts to approximations. In the world of tensor
networks this is known as the problem of approximate
contraction, whereas in the world of graphical models this is known
as the problem of approximated inference and local marginals. 

Over the years many different algorithms and techniques have been
suggested to this problem for both frameworks. Some of them have
been adopted and adjusted to the other framework. For example, the
corner transfer method (CTMRG)\cc{ref:Orus2009:CTMRG,
ref:Nishino1996-CTMRG} for approximate tensor-network contraction
has its roots in Baxter's work in statistical
mechanics\cc{ref:Baxter1968-CTM1, ref:Baxter1978-CTM2}, as well as
ideas of using Monte-Carlo sampling for TN
contraction\cc{ref:Sandvik2007-QC-TN, ref:Wang2011-MC-TN,
ref:Ferris2012-MC-TN}. From the other side, the Tensor
Renormalization Group algorithm (TRG) for the contraction of tensor
networks can be used for highly accurate approximations of classical
statistic mechanical quantities such as partition functions and
magnetization\cc{ref:Levin2007-TRG} (see also
\Ref{ref:Efrati2014-TRG}).

In this paper we show how an important class of inference and
marginalization algorithms for graphical models, called Belief
Propagation (BP) algorithms, can be adapted and used for approximate
tensor network contraction. This idea  was suggested in
\Ref{ref:Robeva2017-TN-Duality}, in the context of a general mapping
between tensor-networks and graphical models. Here, by using a
slightly different mapping, we show how this approximation is in
fact equivalent to the basic contraction approximation that is at
the heart of the simple-update algorithm of tensor networks. As we
discuss later, since the underlying approximation in the BP
algorithm is the Bethe-Peierls approximation, our results imply that
this type of approximation is also at the center of the
simple-update method. It also motivates the study of various
improvements of the BP algorithms as potential tensor-network
contraction algorithms.

\section{A BP algorithm for tensor networks}

Belief Propagation (BP)\cc{ref:Pearl1982-BP} is a statistical
inference algorithm on graphical models, that can also be used to
approximate their marginals\cc{ref:Wainwright2008-GM,
ref:Koller2009-PGM, ref:Mezard2009-Info}. It is also known as the
\emph{sum-product algorithm} in the context of coding
theory\cc{ref:Kschischang2001-sum-product}, and can also be viewed
as an iterative way to solve the Bethe-Peierls equations of
statistical physics\cc{ref:Bethe1935-Approx, ref:Peierls1936-Ising}.

In what follows, we present a BP variant on a PEPS tensor-network.
For an alternative approach, which first maps the PEPS to a
graphical model, and then uses the BP on that graphical model,
please see \App{sec:DEFG}.

\begin{figure}
  \begin{center}
    \includegraphics[scale=0.75]{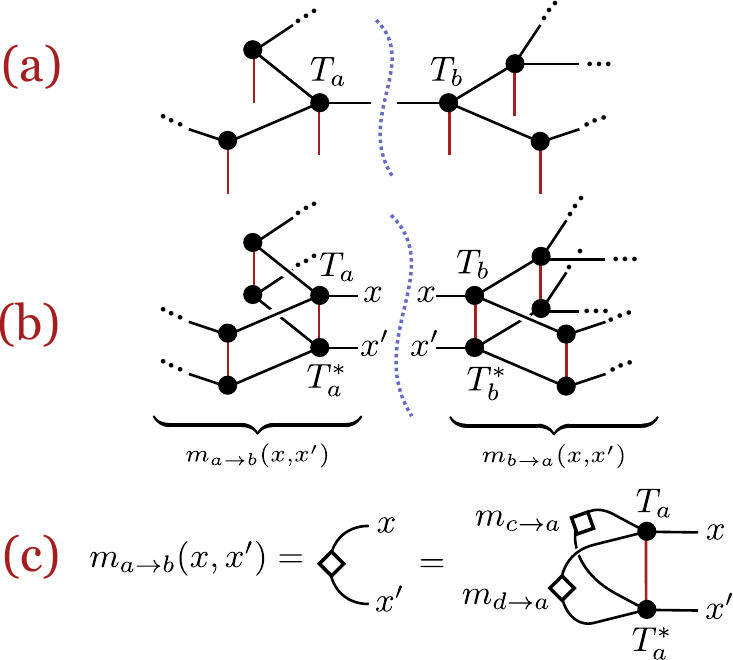}
  \end{center}
  \caption{BP messages on a tree PEPS TN. (a) A local patch of a
  PEPS defined on a tree. The $(a,b)$ edge defines a bi-partition of
  the system into two branches. (b) The TN that corresponds to
  $\braket{\psi}{\psi}$. Also here the $(a,b)$ edge defines two
  branches. The tensors that result from the contraction of each
  branch are the messages $m_{a\to b}(x,x')$ and $m_{b\to a}(x,x')$.
  (c) The messages satisfy a recursion relation, which are used to
  define the BP equations~\eqref{eq:TN-BP}.}
  \label{fig:Tree-TN-messages}
\end{figure}

We consider a PEPS $\ket{\psi}$, in which the physical particles sit
on the vertices (nodes) of some graph $G=(V,E)$
(\Fig{fig:Tree-TN-messages}a). Each node $a\in V$ is associated with
a tensor $T_a$ that has one physical index (leg) of bond dimension
$d$ and an index of dimension $D$ for each adjacent edge, which is
also called a `virtual leg'. Virtual legs of the same edge in $G$
are contracted together. 

Consider a double layer TN that corresponds to the scalar
$\norm{\psi}^2=\braket{\psi}{\psi}$ (\Fig{fig:Tree-TN-messages}b). 
When $G=(V,E)$ is a tree (like an MPS, for example), it can be
contracted efficiently using dynamical programming.  One way to
preform it is as follows. Given two incident nodes, $a,b$, the edge
that connects them divides the system into two parts, see
\Fig{fig:Tree-TN-messages}a,b. We define the ``message'' $m_{a\to
b}(x,x')$ to be the tensor that results from contracting the
branch of $\braket{\psi}{\psi}$ that is connected to $a$, where
$(x,x')$ are the indices of the open ket-bra edges connected to node
$a$. Similarly, the message $m_{b\to a}(x,x')$ is related to the
contraction over the branch of $b$. See
\Fig{fig:Tree-TN-messages}a,b. If we consider them as matrices of
the indices $(x,x')$, they are positive semi-definite, due to the
fact that they are a result of a contraction of a branch with its
complex conjugate. Crucially, these messages satisfy a recursive
relation. If $N_a$ is the set of nodes that are incident to $a$,
then the tensor $m_{a\to b}$ is given by the contraction
\begin{align}
\label{eq:TN-messages}
  m_{a\to b} = \Tr \Big( T_a T_a^* \prod_{a'\in N_a\setminus\{b\}}
    m_{a'\to a}\Big)
\end{align}
where $\Tr(\cdot)$ denotes contraction of joint indices. For
example, If $b,c,d$ are incident to $a$, then the message $m_{a\to
b}(x,x')$ is given in terms of the messages $m_{c\to a}(x,x')$ and
$m_{d\to a}(x,x')$, as shown in \Fig{fig:Tree-TN-messages}c. 

\begin{figure}
  \begin{center}
    \includegraphics[scale=0.75]{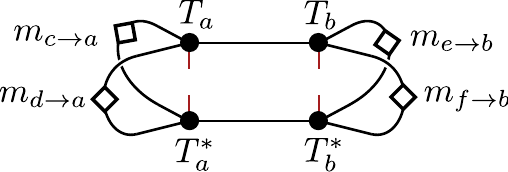}
  \end{center}
  \caption{In a tree, the converged BP messages can be used to
  calculated the local RDMs. In this example, a 2-local RDM. The
  same formulas are used to approximate the local RDMs also when
  the underlying PEPS is not a tree.}
  \label{fig:BP-RDM}
\end{figure}

In principle, we can pick any node which is not a leaf, define it as
a root, and use \Eq{eq:TN-messages} to calculate the messages from
the leaves to the root. Using the messages that lead to the root, we
can calculate $\norm{\psi}^2$. This calculation can also be done
differently. Instead of forcing a particular causality order between
the messages, we can try to solve \Eq{eq:TN-messages} for \emph{all}
messages simultaneously. This can be done by solving
\Eq{eq:TN-messages} iteratively: starting from a set of random 
positive semi-definite (PSD) messages $\{m^{(0)}_{a\to b}(x,x')\}$
for all incident nodes $a,b$, we define the set of messages at step
$t+1$ using the messages of step $t$:
\begin{align}
\label{eq:TN-BP}
  m^{(t+1)}_{a\to b} \EqDef \Tr \Big( T_a T_a^* 
    \prod_{a'\in N_a\setminus\{b\}}  m^{(t)}_{a'\to a}\Big) .
\end{align}
Equation~\eqref{eq:TN-BP} is the BP equation for PEPS
tensor-network. It is a natural extension of the BP equations for
graphical models\cc{ref:Wainwright2008-GM, ref:Koller2009-PGM,
ref:Mezard2009-Info}.  The equation also guarantees that if the
messages at $t$ are PSD when viewed as a
matrix whose $(x,x')$ element is $m_{a\to b}(x,x')$, then so would
the messages at $t+1$ be. The fixed point of this iterative process
will solve \Eq{eq:TN-messages}, and give us all the $m_{a\to
b}(x,x')$ messages. It is a well-known fact that the BP iterations
on tree graphical models have a unique fixed point to which they
converge in linear number of steps\cc{ref:linear-time}. The same
arguments easily generalizes also to our case. Once we have the
messages, we can use the fact that they are contractions over
branches, and use them to calculate local reduced density matrices
(RDM). For example, the calculation of 2-local RDMs are shown in
\Fig{fig:BP-RDM}. The PSD property of the messages guarantees the
resultant RDMs are also PSD.

Thus far, the BP equations~\eqref{eq:TN-BP} might seem no more than
an elegant method for contracting \emph{tree} PEPS. Things become
interesting when we consider graphs with loops. In such case, the
messages can no longer be defined as contraction of branches, since
an edge in the graph no longer partitions it into two distinct
branches. Yet we can still define them as solutions to
\Eq{eq:TN-messages}, and try to find them iteratively using
\Eq{eq:TN-BP}. If the iterations converge to a fixed point, we can
use the messages to estimate local marginals, using the \emph{same}
expressions as we did in the case of trees, which are illustrated in
\Fig{fig:BP-RDM}. This procedure is called \emph{loopy BP}.
Evidently, this is an uncontrolled approximation. In fact, there is
no guarantee the BP iterations will converge to a fixed point, or
that there is a unique fixed point. Nevertheless, in the world of
graphical models, the BP method often provides surprisingly good
results, in particular when the graph looks locally like a tree and
there are no long-range correlations. A famous example are problems
related to the decoding of error correction codes, in which BP
performs extremely well\cc{ref:McEliece1998-BP}. As we shall see,
also in the world of tensor-networks loopy BP often performs well on
problems with short-range entanglement.

While a general theory to explain the performance of the BP
algorithm is still lacking, there are some partial results in this
direction. An important result is due to Yedidia
\etal\cc{ref:Yedidia2001-GBP}, who established the correspondence
between fixed points of the BP algorithm and the Bethe-Peierls
approximation\cc{ref:Bethe1935-Approx}. The Bethe-Peierls
approximation is an approximation scheme for classical statistical
mechanics, in which one treats the system as if its defined on a
tree (a Bethe lattice). This assumption implies that the Gibbs
distribution of the system, as well as the free-energy functional
can be written as functions of the local marginals. When the actual
interaction graph of the system is not a tree, this is an
uncontrolled approximation. Nevertheless, also in these cases, one
can take the Bethe free-energy functional and look for locally
consistent set of marginals that minimizes it. This procedure often
gives surprisingly good approximations. Yedidia
\etal\cc{ref:Yedidia2001-GBP} showed that there is a 1-to-1
correspondence between extermum points of the Bethe free-energy and
fixed points of the BP equations. The marginals obtained from the
messages at a fixed point minimizes the Bethe free-energy, and
conversely, from the marginals at the extremum point one can derive
fixed-point messages of BP. 

As we show in \App{sec:DEFG}, this result naturally generalizes to
our case. The idea is that our BP algorithm for PEPS is the usual BP
algorithm that is applied to a graphical model with complex entries,
obtained from the tensor-network of $\braket{\psi}{\psi}$. Such
graphical models were studied in \Ref{ref:Cao2017-DEFG}, under the
name Double Edge Factor Graph (DEFG), where it was argued that the
BP algorithm corresponds, like in the usual case, to the extreme
points of the Bethe free-energy. 

At this point it seems tempting to benchmark the BP algorithm for
PEPS contraction, and compare it to other methods. Indeed, we first
used the BP algorithm as a contraction subroutine in an imaginary
time evolution algorithm on various 2D models, and compared to the
performance of the simple update method\cc{ref:Alkabetz2020-BPU}.
Surprisingly, the final energies of both algorithms were
suspiciously close to each other. As we show next, this can be
explained theoretically; the mean field approximation at the heart
of the simple-update method and the BP algorithm are equivalent.

\section{The simple update method}

The simple-update method\cc{ref:Jiang2008-SU} is a direct
generalization of the TEBD\cc{ref:Vidal2003-TEBD,
ref:Vidal2004-TEBD, ref:Daley2004-TEBD} algorithm for 1D real and
imaginary times evolution to higher dimensions. It calculates the
dynamics of many-body spin systems that sit on a lattice, and are
described by a (quasi-) canonical PEPS tensor-network.  The method
is efficient and numerically stable, but often results in poor
accuracy due to its oversimplified representation of local
environments.

\begin{figure}[h]
  \begin{center}
    \includegraphics[scale=0.75]{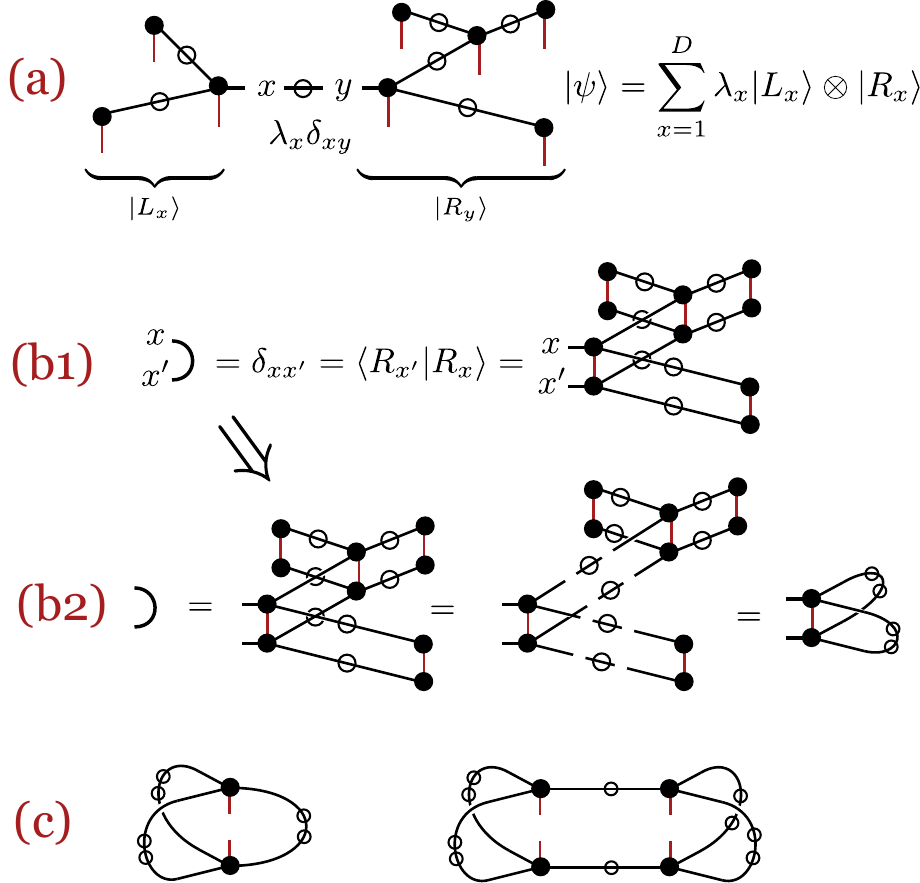}
  \end{center}
  \caption{ The properties of a canonical representation of a tree
    PEPS. (a) An example of a canonical representation of a tree
    PEPS and its relation to the Schmidt decomposition. The empty
    circles are diagonal tensors that correspond to the Schmidt
    weights $\lambda$. (b1) The orthonormality of the Schmidt bases implies
     a simple formula for the contraction of the left and right
    branches and (b2) a \emph{local} canonical condition on the PEPS
    tensors. (c) A local expression for the reduced density
    matrices. } \label{fig:SU-on-tree}
\end{figure}

At its core lies a quasi-canonical form of the PEPS that allows a
crude approximation of local TN environments. This approximation is
often referred to as mean field approximation. When the PEPS is in
the shape of a tree, this form is truly canonical: it corresponds to
consecutive Schmidt decompositions of the many-body quantum state.
Every edge in the graph connects two disjoint branches of the
system, which correspond to the Schmidt decomposition between these
two branches $\ket{\psi}=\sum_{i=1}^D \lambda_i
\ket{L_i}\otimes\ket{R_i}$. Specifically, the TN consists of two
types of tensors: tensors $T_a$ that are connected to the physical
legs, and diagonal $\lambda$ tensors that sit in the middle of every
edge and correspond to the Schmidt weights, as illustrated in
\Fig{fig:SU-on-tree}. The orthonormality of the Schmidt
decomposition, and its correspondence with the TN structure implies
that contraction over branches of the tree is given by a simple
Kronecker delta function (\Fig{fig:SU-on-tree}b). Therefore, the
RDMs of a given region can be readily computed using only the
tensors that surround it (\Fig{fig:SU-on-tree}c), and in addition
the tensors must satisfy \emph{local} canonical conditions shown in
(\Fig{fig:SU-on-tree}b). This type of approximation is often
referred to as \emph{mean field approximation of the environment} in
the TN literature. 

When the underlying graph has loops, the canonical form is no longer
well-defined; removing an edge from the graph no longer divides it
into two parts, and so it cannot be associated with a Schmidt
decomposition between two branches. Nevertheless, we can still
define a PEPS to be quasi-canonical if it satisfies the \emph{local}
canonical constraints of \Fig{fig:SU-on-tree}b2.  In such case, the
expression for local RDMs (e.g., \Fig{fig:SU-on-tree}c) is no longer
exact. Nevertheless, when the graph looks locally like a tree, or
when the quantum state has only short-range correlations, this
approximation is often reasonable.

\begin{figure}[h]
  \begin{center}
    \includegraphics[scale=0.75]{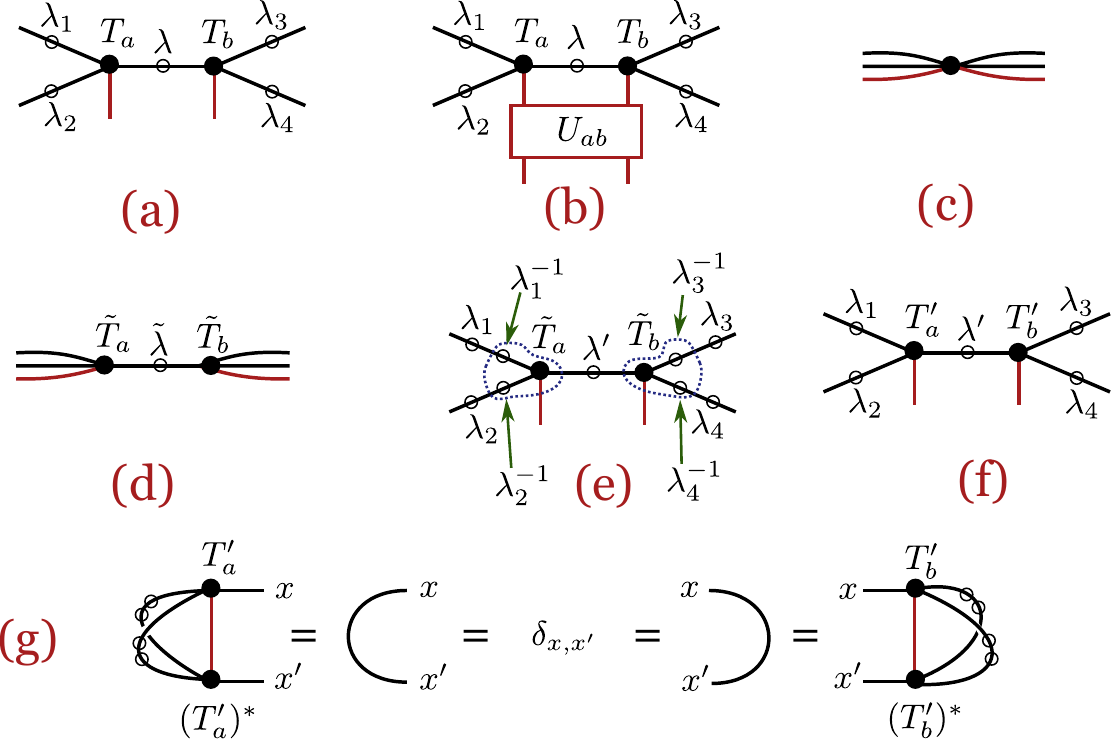}
  \end{center}
  \caption{    
    The Simple-Update steps of applying a ``gate'' $U_{ij}$ on the
    tensors $\{T_i, \lambda, T_j\}$, and updating them to $\{T'_i,
    \lambda', T'_j\}$. (a) The original tensors. (b) Applying
    $U_{ij}$. (c) The $T_a,T_b$ tensors and all surrounding
    $\lambda$ weights are contracted into one big tensor, which is
    reshaped as a matrix. (d) An SVD  is performed on
    the matrix (e) and (f) a trivial $\lambda_i\lambda_i^{-1}$ pairs
    are inserted to the external legs and define the new
    $T_a',\lambda', T_b'$ tensors. At this step, one can
    truncate the smallest weights of $\lambda'$ to reduce the bond
    dimension back to $D$. (g) If no truncation was done, the
    resulting tensors satisfy some of the local canonical conditions. }
    \label{fig:SU-step}
\end{figure}

In the simple-update algorithm one usually starts from a
quasi-canonical PEPS and then applies local gates that performs real
or imaginary time evolution according the Trotter-Suzuki
decomposition. For example, in the imaginary time evolution, if the
Hamiltonian interaction term between the neighboring sites $a,b$ is
$h_{ab}$, the operator $U_{ab} = e^{-\delta\tau h_{ab}}$ will be
applied, where $\delta \tau$ is a small Trotter-Suzuki time step.
Once $U_{ab}$ is applied, a local SVD decomposition is performed, as
shown in \Fig{fig:SU-step}b-f, which guarantees that: (i) the
resultant tensor network can be reshaped into a local PEPS with
$T_a, \lambda, T_b$ replaced by $T_a', \lambda', T_b'$ (ii) a
truncation is performed so that the bond dimension of new tensors
does not increase, and (iii) \emph{some} of the local canonical
conditions are (approximately) satisfied --- see \Fig{fig:SU-step}g.

When the operator $U_{ab}$ is not unitary (e.g., in the case ofs
imaginary time evolution), or when truncation is performed, the
resultant TN will no longer be quasi-canonical. Some of the
canonical conditions will be satisfied, not all of them. However,
also in that case, the $\lambda$ weights still provide a reasonable
approximation for the local environments, as is evident by the
success of the SU algorithm in many cases. In particular, in the
imaginary time case, if the system reaches a fixed point, (the
approximate ground state), it is also a fixed point of all the local
SU steps. This state satisfies \emph{all} the local canonical
conditions, and is therefore a quasi-canonical state.

We conclude this section by noting that the SU steps can be easily
turned into an algorithm for finding a quasi-canonical form of a
PEPS TN. Starting from an arbitrary PEPS, we can apply a ``trivial''
SU step with $U_{ab}=\Id$, without the truncation step. In other
words, we essentially perform an SVD decomposition of the fused
tensor in \Fig{fig:SU-step}d, which yields tensor $T_a', \lambda',
T_b'$ satisfying the local canonical condition of
\Fig{fig:SU-step}g. Repeating this step on all edges, if a fixed
point is reached, it is by definition a quasi-canonical PEPS, and
can be used to calculate local RDMs via the $\lambda$ weights (see
\Fig{fig:SU-on-tree}c). We call this algorithm \emph{trivial-SU},
and note that it has already been used previously (see, e.g., the
quasi-orthogonalization in Sec.~B of \Ref{ref:Kalis2012-SU}, or the
super-orthogonalization of \Ref{ref:Ran2012-Tucker}). See also
interesting parallels between our BP construction, the trivial-SU
and the NCD theory of \Ref{ref:Ran2013-NCD} (see also Chapter~5 in
\Ref{ref:Ran2020-TN}).

\section{BP-SU equivalence}

The trivial-SU algorithm and the BP algorithm for TN are two
different algorithms for approximate contraction of PEPS. They
originate from two very different places: the trivial-SU algorithm
is a natural algorithm for TN, which relies on the Schmidt
decomposition, whereas the BP algorithm is a message-passing
algorithm for graphical models that originated from inference
problems and the Bethe-Peierls approximation. It might therefore
come as a surprise that these two algorithms are equivalent. In
hindsight, this could have been anticipated, as both are iterative
algorithms that are exact on trees. We prove
\begin{theorem}
\label{thm:duality} Every trivial SU fixed point corresponds to a BP
  fixed point such that the local RDMs computed in both methods are
  identical.
\end{theorem}
As a simple corollary of this theorem we conclude that if the
trivial SU equations and the BP equations have a unique fixed point,
both algorithms will yield the same RDMs. In light of this
equivalence, the success of the imaginary time SU algorithm for many
models (see, for example, models analyzed in
\Ref{ref:Jahromi2019-SU}), is another example of the success of the
Bethe-Peierls approximation.

To prove \Thm{thm:duality}, we note that in the BP algorithm the TN
remains fixed, while the BP messages evolve to a fixed point. In the
trivial-SU, there are no messages, but the local tensors that make
up the TN evolve until they converge to a quasi-canonical
fixed-point, without changing the underlying quantum state. In both
cases, the evolution is done via local steps. Our proof uses two
lemma:
\begin{lemma}
\label{lem:Lemma-I} Let $\mcT, \mcT'$ be two tensor-networks that
  represent the same state $\ket{\psi}$, such that $\mcT'$ is
  obtained from $\mcT$ using a single trivial SU step on tensors
  $T_a, T_b$ and the $\lambda$ weight between them. Then every BP
  fixed-point of $\mcT$ has a corresponding fixed-point of $\mcT'$
  with the same RDMs, and vice-versa.
\end{lemma}
The idea of the proof is to show that the fixed point messages of
the new TN can be constructed from the fixed point messages
of the old TN, except for the local place of change, where the
messages are adapted to fit the new tensors. The full proof of the
lemma is given in \App{sec:proofs}.

Using this lemma repeatably along the trivial-SU iterations, we
conclude that the BP fixed points of an initial TN is equivalent to
those of its quasi-canonical representation. To finish the proof, we
show that the BP fixed point of a quasi-canonical PEPS yields the
same RDMs as the $\lambda$ weights do:
\begin{lemma}
  \label{lem:Lemma-II} Given a TN in a quasi-canonical form (i.e., a
  fixed point of the trivial SU algorithm), it has a BP fixed point
  that gives the same RDMs estimates as those of the quasi-canonical
  form based on the $\lambda$ weights.
\end{lemma}
The idea of the proof is that after ``swallowing'' a
$\sqrt{\lambda}$ of each $\lambda$ tensor in its two adjacent
$T_a,T_b$ tensors, we reach a PEPS for which the messages $m_{a\to
b}(x,x') = \lambda_x \delta_{x,x'}$ are a BP fixed-point. The full
proof is found in \App{sec:proofs}.

\section{Numerical tests.}

In this section we present some numerical tests that compare the BP
method to the trivial simple-update method.  The asymptotic
complexity of one step in both algorithms is similar. For example,
on a square grid with virtual bond dimension $D$ and physical bond
dimension $d$, both steps take $O(dD^5)$ basic arithmetic
operations.  What might be more interesting is the number of
iterations needed for convergence, which can be tested numerically.
We compared these numbers on two types of PEPS on finite square
grids. One type was a random PEPS with normal complex entries and the
other was an approximate ground state of the anti-ferromagnetic
Heisenberg model on a square lattice with random, nearest-neighbors
coupling. For every model we ran the BP algorithm and the trivial-SU
algorithm on $20-50$ random instances to calculate the 2-body RDMs.
For each instance we calculated the ratio of the convergence time
(i.e., number of iterations until convergence) $T_{BP}/T_{tSU}$. The
results are given in Table~\ref{tab:statistics}. While for the
random PEPS instances the BP method seems slightly faster, in
the AFH models, where correlations are of longer range, the
trivial-SU seems to converge faster, in particular as the bond
dimension $D$ increases. The full histogram of the results, as well
as the full numerical details can be found in
\App{sec:extra-numerics}.

\begin{table}
  \begin{tabular}{rccccc}
     & \multicolumn{2}{c}{Random-PEPS} &\quad& \multicolumn{2}{c}{AFH}\\
    &$4\times 4$ & $10\times 10$ &\quad& $4\times 4$ & $10\times 10$ \\
    \hline
    $D=2$: & $0.7\pm 0.2$ & $0.6\pm 0.2$ &\quad& 
             $1.4\pm 0.1$ & $1.3\pm 0.1$\\
    $D=3$: & $0.7\pm 0.2$ & $0.7\pm 0.2$ &\quad& 
             $1.1\pm 0.2$ & $1.2\pm 0.1$\\
    $D=4$: & $0.7\pm 0.2$ & $0.8\pm 0.2$ &\quad& 
             $1.6\pm 0.5$   & $1.8\pm 0.2$\\
  \end{tabular}
  \caption{Average ratio of convergence times $T_{BP}/T_{tSU}$,
    together with standard deviations for random
    PEPS and ground states of anti-ferromagnetic Heisenberg model
    (AFH) with random n.n. couplings on a $N\times N$ lattices. We
    simulated different bond dimensions $D=2,3,4$ using $20-50$
    realizations for each configuration. Full details can be found
    in \App{sec:extra-numerics}.} 
  \label{tab:statistics}
\end{table}

\section{Discussion}

In this work we established a bridge between the world
of graphical models and tensor networks. We have defined the Belief
Propagation method for PEPS contraction, which can be viewed as the
ordinary BP method applied for Double Edge Factor Graphs
(DEFG)\cc{ref:Cao2017-DEFG} that are derived from the PEPS tensor
network. Just as in the ordinary graphical models case, the fixed
points of the BP iterations correspond to extreme points of the
Bethe free-energy, which is defined for the underlying PEPS TN. We
have shown that the BP algorithm is equivalent to the trivial SU
algorithm on PEPS tensor networks, which leads to a quasi-canonical
form. This correspondence has some interesting implications. First,
since the fixed points of the imaginary time SU algorithm is a
quasi-canonical PEPS, our result implies that its SU approximate
environments correspond to a Bethe-Peierls approximation. The
success of the SU algorithm can therefore be seen as another example
of the power of the Bethe-Peierls approximation. Second, it shows
that one of the simplest algorithms for approximating marginals in
the world of graphical models is equivalent to one of the simplest
approximate contraction algorithms in the world of tensor networks.

The equivalence of these two methods, which come from different
fields and are derived from different principles, is interesting for
several reasons. From the practical point of view, we can try to
``import'' other, more sophisticated algorithms for marginal
approximations to the world of tensor networks. A natural candidate
is the Generalized Belief-Propagation (GBP)
algorithm\cc{ref:Yedidia2001-GBP}, which generalizes the BP
algorithm by considering messages from larger regions in the graph.
Just as the BP algorithm converges to extreme points of the Bethe
free-energy, the GBP algorithm converges to extreme points of the
Kikuchi free-energy of the cluster variation
method\cc{ref:Kikuchi1951-CVM, ref:Morita1994-CVM}. As shown in
\Ref{ref:Yedidia2001-GBP}, this algorithm provides a much better
approximation of the marginals, at the price of a higher
computational cost. It would be interesting to compare the
performance of this algorithm, as well as other BP
improvements\cc{ref:Montanari2005-LCB, ref:Chertkov2006-LCB, 
ref:Mooij2007-LCB, ref:Nachmani2016-NeuralBP,
ref:Cantwell2019-Loops}, when acting on TNs to that of more accurate
contraction algorithms, such as the corner transfer method
(CTMRG)\cc{ref:Orus2009:CTMRG, ref:Nishino1996-CTMRG}, the
tensor-network renormalization group (TRG)\cc{ref:Levin2007-TRG,
ref:Gu2008-TRG} and boundary MPS (bMPS)\cc{ref:Jordan2008-bMPS}, to
name a few. 

From the theoretical prospective, it would be interesting to better
understand the physical and mathematical role of the complex Bethe
free-energy that we have derived. In particular, we know that for
tree-tensor networks, it is related to the Schmidt decomposition.
Can we somehow relate it to the underlying entanglement
structure also when the underlying graph has loops? Another
interesting question is whether the BP equations can be used to
analytically analyze models for which the ground state is a known
PEPS, such as the AKLT model.

\section{Acknowledgments} 

The authors thank Eyal Bairy, Raz Firanko, Roman Or\'us and Yosi
Avron for their help with the manuscript and many useful
suggestions. IA acknowledges the support of the Israel Science
Foundation (ISF) under the Individual Research Grant No.~1778/17.

\bibliographystyle{apsrev4-1}
\bibliography{BP-and-TN}

\onecolumngrid
\appendix

\section{Graphical models, Belief Propagation and the mapping of 
  PEPS tensor networks and double-edge factor graphs}
\label{sec:DEFG}

In this section we give a very brief background to the subject of
graphical models and the Belief Propagation algorithm, and then
sketch a mapping between the PEPS tensor networks to a particular
type of graphical models called double-edge factor graphs. Together,
this will show that our BP algorithm for PEPS is essentially the
usual BP algorithm applied to the mapped double-edge factor graphs,
and the converged messages correspond to a Bethe-Peierls
approximation. 

\subsection{Graphical models and Belief Propagation}

Graphical models are a powerful tool for modeling multivariate
probability distributions.  They provide a succinct description of
the statistical dependence of a set of random variables using
graphs, and are used in fields such as bioinformatics, communication
theory, statistical physics, combinatorial optimization, signal and
image processing, and statistical machine learning to name a few. In
this section we give a brief background on this tool and the BP
algorithm. For a thorough review, we refer the reader to
Refs.~\cite{ref:Wainwright2008-GM, ref:Koller2009-PGM,
ref:Mezard2009-Info}.

Roughly speaking, there are three main families of graphical models:
Bayesian networks, Markov Random Fields (MRF), and factor graph
graphical models. As the latter family supersedes the first two, we
will concentrate on it.

A factor graph graphical model is a succinct description of 
probability distribution
$P(x_1,\ldots, x_n)$ of a set of random variables $X_1, X_2, \ldots
X_n$. It is given as a product
\begin{align*}
  P(x_1,\ldots, x_n) 
    = \frac{1}{Z}\prod_{a \in \mcF}  f_a(\Bx_a),
\end{align*}
where $\mcF$ is a collection of ``factors'' $f_a(\Bx_a)$. These are
non-negative functions of  small  subsets of variables $\Bx_a = (x_{i_1},
x_{i_2}, \ldots, x_{i_k})$. $Z\EqDef\sum_{\Bx}\prod_{a \in \mcF}
f_a(\Bx_a)$ is an overall normalization constant. A very natural
distribution of that form appears in classical statistical mechanics.
Given a local Hamiltonian $H(\Bx) = \sum_{a\in \mcF} h_a(\Bx_a)$,
its Gibbs distribution is 
\begin{align*}
  P(\Bx) =\frac{1}{Z}e^{-\beta H(\Bx)} = \frac{1}{Z}
    \prod_{a\in \mcF} e^{-\beta h_a(\Bx_a)}.
\end{align*}
In this case $f_a(\Bx_a)=e^{-\beta h_a(\Bx_a)}$
and $Z$ is the partition function.

\begin{figure}
  \begin{center}
    \includegraphics[scale=1]{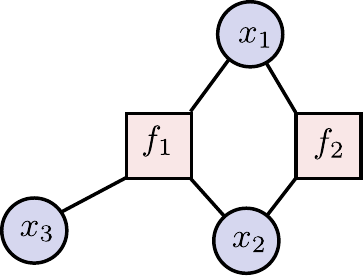}
  \end{center}
  \caption{An example of a factor graph representing the probability
  distribution $P(x_1, x_2, x_3) = \frac{1}{Z} f_1(x_1, x_2,
  x_3)\cdot f_2(x_1, x_2)$.}
  \label{fig:factor-graph}
\end{figure}

There is convenient graphical way to capture the relation between
the various factors, using a so-called \emph{factor graph}
$\mcG=(\mcV, \mcF, \mcE)$. This is a bipartite graph with two types
of vertices: $\mcV=\{x_1, x_2, \ldots, x_n\}$ is the set of
variables, also called \emph{nodes}.  The other set of vertices are
the factors $\mcF=\{f_1, f_2, \ldots\}$. $\mcE$ is the set of edges,
where an edge connects the node $x_i$ to the factor $f_a$ iff $f_a$
depends on $x_i$. For example, the factor graph in
\Fig{fig:factor-graph} corresponds to probability distributions of
the form $P(x_1, x_2, x_3) = \frac{1}{Z}f_1(x_1,x_2,x_3)\cdot
f_2(x_1,x_2)$.

Given a graphical model, a central task is to calculate its marginal
over some small set of random variables. This is needed, for
example, for the calculation of local expectation values, or for the
optimization of the model with respect to empirical data. This task
is NP-hard in general, involving a summation over an exponential
number of configurations\cc{ref:Cooper1990-BPcomp}.

Belief Propagation (BP)\cc{ref:Pearl1988-BP, ref:Wainwright2008-GM,
ref:Koller2009-PGM, ref:Mezard2009-Info} is a message-passing
algorithm that is designed to approximate such marginals. It is
exact on graphical models whose underlying graph is a tree and often
gives surprisingly good results on loopy graphs. In these cases,
however, it is essentially an uncontrolled heuristic. The BP
algorithm is often known in different names at different contexts.
In statistical physics, it is known as the `Bethe–Peierls
approximation'\cc{ref:Bethe1935-Approx, ref:Peierls1936-Ising}, and
in coding theory as `sum-product
algorithm'\cc{ref:Kschischang2001-sum-product}. The name `belief
propagation' was coined by J.~Pearl, who used it in the context of
Bayesian networks\cc{ref:Pearl1988-BP, ref:Pearl1982-BP}.

\begin{figure}
  \begin{center}
    \includegraphics[scale=0.75]{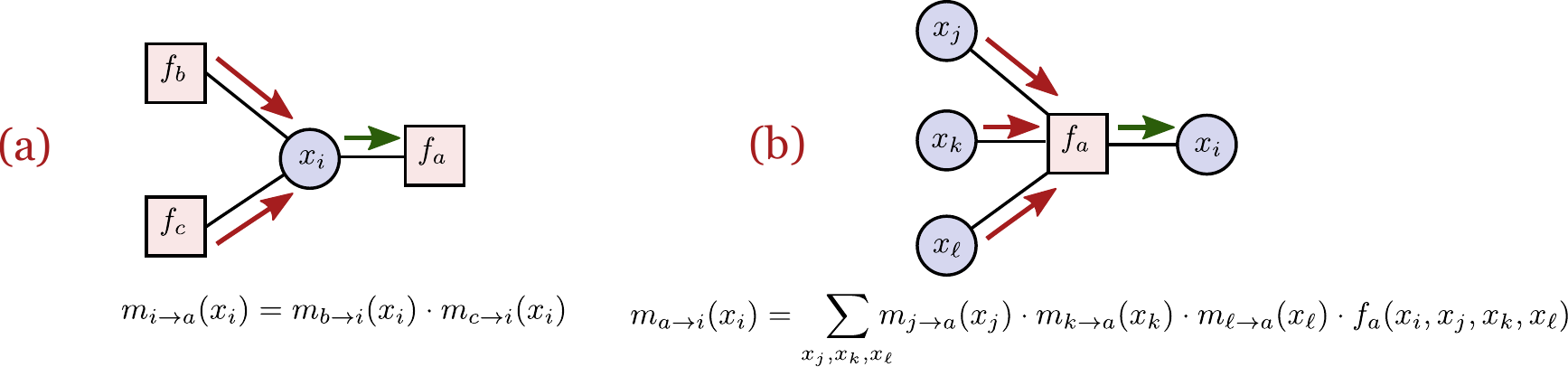}
  \end{center}
  \caption{Illustration of the BP equations on a factor-graph
    graphical model ---
    Eqs.~(\ref{eq:BP-i-to-alpha},\ref{eq:BP-alpha-to-i}). In such
    case there are two types of messages: nodes
    to factors (Fig. a, \Eq{eq:BP-i-to-alpha}) and factors to nodes
    (Fig. b, \Eq{eq:BP-alpha-to-i}).}
  \label{fig:BP-equations}
\end{figure}

The main objects in the BP algorithm are ``messages'' between factor
and nodes and vice versa. To write them, let us define $N_i$ as the
set of factors in which $x_i$ participates, and similarly $N_a$ to
be the set of variables of the factor $f_a$ (i.e., adjacent nodes to
$f_a$). A message from a factor $f_a$ to an adjacent node $i\in N_a$
is a non-negative function $m_{a\to i}(x_i)$ and a message from node
$i$ to factor $a$ is a non-negative function $m_{i\to a}(x_i)$. The
BP algorithm starts by initializing the messages (say, randomly),
and then at each step, the messages are updated from the messages of
the previous step by the local rules (see also \Fig{fig:BP-equations}):
\begin{align}
\label{eq:BP-i-to-alpha}
  m^{(t+1)}_{i\to a}(x_i) &\EqDef \prod_{b\in
    N_i\setminus\{a\}} m^{(t)}_{b\to i}(x_i) , \\
  m^{(t+1)}_{a\to i}(x_i) &\EqDef
  \sum_{\Bx_a\setminus\{x_i\}} f_a(\Bx_a)
    \prod_{j\in a\setminus\{i\}} m^{(t)}_{j\to a}(x_j) .
\label{eq:BP-alpha-to-i}
\end{align}

\begin{figure}
  \begin{center}
    \includegraphics[scale=0.75]{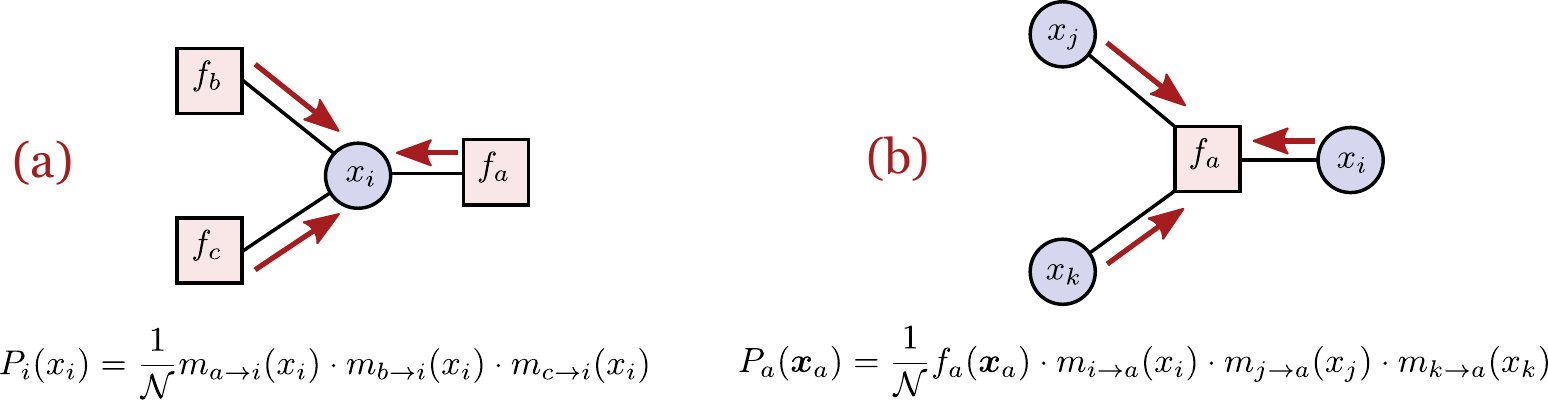}
  \end{center}
  \caption{Calculating the local marginals from the BP messages by
    the formulas in Eqs.~(\ref{eq:marginal-i},\ref{eq:marginal-a}).
    These formulas give the exact marginals when the underlying
    graphical model is a tree.} \label{fig:BP-marginals}
\end{figure}

If the messages converge to a fixed point, they can be used to
estimate the marginals on subsets of nodes. For example, the
marginal of a single variable $x_i$ is given by
\begin{align}
\label{eq:marginal-i}
  P_i(x_i) = \frac{1}{\mcN} \prod_{a\in N_i} m_{a\to i}(x_i) ,
\end{align}
where $\mcN$ is a normalization factor. The 
marginal over the variables of a factor are given by
\begin{align}
\label{eq:marginal-a}
  P_a(\Bx_a) = \frac{1}{\mcN} f_a(\Bx_a) \prod_{i\in N_a} m_{i\to a}(x_i) ,
\end{align}
These two expressions are demonstrated in \Fig{fig:BP-marginals}

For tree graphical models, the BP messages are promised to converge
to a unique fixed point in linear time, and 
formulas~(\ref{eq:marginal-i}, \ref{eq:marginal-a}) give the exact
marginals\cc{ref:linear-time}.

When the underlying graph has loops, the BP algorithm is called
`\emph{loopy BP}', and the formulas for the marginals become,
essentially, uncontrolled. Moreover, it is not known how fast the
algorithm will converge, if ever, or if it has a unique fixed point.
Nevertheless, in many practical cases, the loopy BP provides
surprisingly good result. 

While, a general theory to explain the performance of BP algorithm
is still lacking, there are some partial results in this direction.
An important result is due to Yedidia \etal\cc{ref:Yedidia2001-GBP},
who highlighted the correspondence between the Bethe-Peierls
approximation and fixed points of the BP algorithm, which we now
explain briefly. The starting point are 
models defined on tree graphs. A simple observation is that for
these models, the global probability distribution can be written in
terms of its local marginals:
\begin{align}
\label{eq:tree-P}
  P(\Bx) = \prod_{a\in\mcF} P_a(\Bx_a)
    \prod_{i\in \mcV} \big(P_i(x_i)\big)^{1-d_i} ,
\end{align}
where $P_a(\Bx_a)$ is the marginal on the nodes adjacent to
$a\in\mcF$, and $P_i(x_i)$ is the marginal of $x_i$. Finally,
$d_i=|N_i|$ is the number of factors that are adjacent to $x_i$. Using
\Eq{eq:tree-P}, we can write the free energy in terms of the local
marginals:
\begin{align}
\label{eq:F-Bethe}
  F_{Bethe} = \sum_a \sum_{\Bx_a}P_a(\Bx_a)
    \ln\frac{P_a(\Bx_a)}{f_a(\Bx_a)}
     - \sum_i(d_i-1)\sum_{x_i}P_i(x_i)\ln P_i(x_i) .
\end{align}
The above expression is called the \emph{Bethe free-energy}. When
the underlying graph is not a tree, the Bethe free-energy is still
well defined, but no longer equal the exact free-energy.  In such
case, we can use it to approximate the local marginals. We write the
Bethe free-energy as a function of unknown marginals $\{q_a(\Bx_a),
q_i(x_i)\}$, and then we estimate the real marginals $\{ P_a(\Bx_a),
P_i(x_i)\}$ by finding the $\{q_a(\Bx_a), q_i(x_i)\}$ that
\emph{minimize} the Bethe free-energy. This procedure is exact on
trees, where the Bethe free-energy is equal to the exact free
energy, but on loopy graphs it is essentially an uncontrolled
approximation; the resultant $q_a(\Bx_a), q_i(x_i)$, may be far from
the actual marginals, and in fact, they might not be marginals of
any underlying global distribution. Nevertheless, decades of
experience in statistical mechanics has shown that this is often a
good approximation that gives better results than simple mean field.
In \Ref{ref:Yedidia2001-GBP} it was showen that there is a
one-to-one connection between the fixed-points of the BP equations
and the extreme points of the Bethe free-energy. The Lagrange
multipliers used to minimize the latter become the fixed-point BP
messages, and the local marginals coincide. This connection between
a message-passing inference algorithm, and a variational approach
gave rise to a plethora of other message-passing algorithms, such as
generalized belief propagation (GBP), which are based on more
sophisticated free energies, such as Kikuchi's cluster variation
method\cc{ref:Kikuchi1951-CVM}

\subsection{Mapping a PEPS tensor network to a graphical model}

In this section we present a mapping that takes a PEPS TN to a
graphical model. Relations and dualities between graphical models
and tensor networks have been studied over the years by several
authors\cc{ref:Critch2014-TN-GM, ref:Robeva2017-TN-Duality,
ref:Chen2018-RBM}. Our approach shares some similarities with these
works, but in particular builds on the Double Edge Factor Graph
(DEFG) formalism of \Ref{ref:Cao2017-DEFG}. This allows us to
transform a tree tensor-network into a tree graphical model, and it
also has the desirable property of messages being positive
semi-definite matrices.

\begin{figure}
  \begin{center}
    \includegraphics[scale=1]{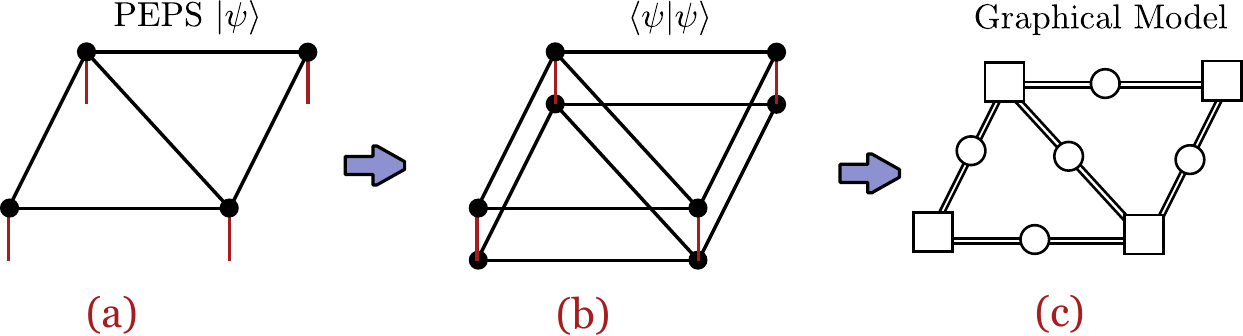}
  \end{center}
  \caption{\label{fig:map-TN-to-GM} Mapping a tensor network to a
  graphical model of type double edge factor graph.}
\end{figure}

\begin{figure}
  \begin{center}
    \includegraphics[scale=0.75]{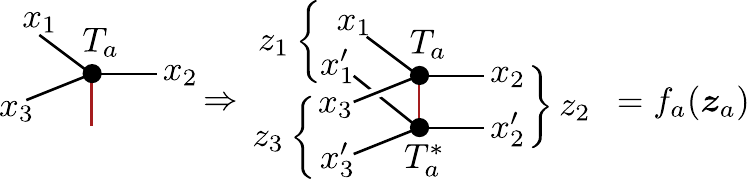}
  \end{center}
  \caption{Defining the complex $f_a(\Bz_a)$ factors from the PEPS
  tensors $T_a$, via $f_a(\Bz_a) \EqDef \Tr(T_a T_a^*)$. $z_i\EqDef
  (x_i, x_i')$ that originate from the ket and the bra of the
  $\braket{\psi}{\psi}$ TN.} \label{fig:Fa}
\end{figure}

The mapping between PEPS and DEFG is illustrated in
\Fig{fig:map-TN-to-GM}. Let $\ket{\psi}$ be the many-body quantum state
that is described by our TN, and consider the tensor network
corresponding to $\braket{\psi}{\psi}$, in which, we clump every
edge in $\ket{\psi}$ with its equivalent edge in $\bra{\psi}$ (see
\Fig{fig:map-TN-to-GM}b). We call such pairs of edges `double
edges'. They run over $D^2$ values of the double indices
$(x,x')$ of the ket and the bra TN. We map this TN into a graphical
model as follows: 
\begin{itemize}
  \item We associate every double edge with a node so that 
    its double indices $(x_i, x_i')$ now become a single variable in
    the graphical model. We denote this pair by a single variable 
    $z_i=(x_i, x_i')$, and notice that it runs over $D^2$ discrete
    values.
    
  \item We associate the contraction of every pair $T_a,T^*_a$ of
    bra-ket local tensors along their physical leg with a factor.
    See \Fig{fig:map-TN-to-GM}b,c. Specifically, let $T^\mu_{a; x_1,
    \ldots, x_k}$ be the PEPS tensor at node $a$ with $\mu$ being
    the physical leg, then the resultant factor is given by 
    \begin{align}
    \label{def:fa}
      f_a[z_1, \ldots, z_k] \EqDef f_a\big[ (x_1, x'_1), 
        \ldots, (x_k, x_k')\big]
        \EqDef \sum_{\mu=1}^d T^\mu_{a; x_1, \ldots, x_k}\cdot
          \Big(T^\mu_{a; x'_1, \ldots, x'_k}\Big)^*
    \end{align}
    See \Fig{fig:Fa}. As in the body of the paper, we write
    $\Bx_a=(x_1, \ldots, x_k)$, $\Bx'_a=(x'_1, \ldots, x'_k)$, and
    $\Bz_a=(z_1, \ldots, z_k)$ for the variables of the factor $a$.
    With this notation, we may write $f_a(\Bz_a) = f_a(\Bx_a,
    \Bx_a')$. Definition~\ref{def:fa} immediately implies that as a
    matrix, $f_a(\Bx_a, \Bx_a')$ is positive semi-definite.

  \item Graphically, variable nodes are denoted by circles, and factors by
    squares. Adjacent variables and factors are connected by double
    lines (edges) that correspond to the double variable
    $z_i=(x_i,x_i')$ that they represent. See \Fig{fig:map-TN-to-GM}.
\end{itemize}
With these definitions, the resultant graphical model is called a
DEFG and describes the function
\begin{align}
\label{def:Pz}
  P(\Bz) &\EqDef P(z_1, \ldots, z_n) = \frac{1}{Z} \prod_a
  f_a(\Bz_a) , & 
    Z &\EqDef \sum_{\Bz} \prod_a  f_a(\Bz_a) = \braket{\psi}{\psi}.
\end{align}
Writing $P(\Bz)$ as $P(\Bx,\Bx')$, the positivity of the individual
$f_a(\Bx_a,\Bx'_a)$ implies that also $P(\Bx,\Bx')$ is a positive
semi-definite function. We can therefore interpret it as the density
matrix of some fictitious quantum states that ``lives on the edges
of the PEPS'', although it has a non-conventional normalization
because $\Tr P = \sum_{\Bx,\Bx}P(\Bx,\Bx)$ is not necessarily equal
to $1$ (instead, it is $\sum_{\Bx,\Bx'}P(\Bx,\Bx')=1$).

Once the factor graphical model is defined, we can run the BP
iterations on it, 
\begin{align}
\label{eq:BP-i-to-alpha-DEFG}
  m^{(t+1)}_{i\to a}(z_i) &\EqDef \prod_{b\in
    N_i\setminus\{a\}} m^{(t)}_{b\to i}(z_i) , \\
  m^{(t+1)}_{a\to i}(z_i) &\EqDef
  \sum_{\Bz_a\setminus\{z_i\}} f_a(\Bz_a)
    \prod_{j\in a\setminus\{i\}} m^{(t)}_{j\to a}(z_j) .
\label{eq:BP-alpha-to-i-DEFG}
\end{align}
which are simply the usual BP equations~(\ref{eq:BP-i-to-alpha},
\ref{eq:BP-alpha-to-i}) with $x_i$ replaced by the double-edge
variable $z_i$. It is easy to see that these equations are
equivalent to \Eq{eq:TN-BP} in the paper by noting that every node
$i$ is adjacent to exactly two factors $a,b$ (because it corresponds
to an edge in the PEPS connecting two vertices), and therefore by
\Eq{eq:BP-i-to-alpha-DEFG},
\begin{align*}
  m^{(t+1)}_{i\to b}(z_i) = m^{(t)}_{a\to i}(z_i),
\end{align*}
which we identify with $m^{(t+1)}_{a\to b}$ from \Eq{eq:TN-BP}.
Moreover, the summation $\sum_{\Bz_a\setminus\{z_i\}}$ in
\Eq{eq:BP-alpha-to-i-DEFG} is exactly the contraction of the virtual
legs in \Eq{eq:TN-BP} and \Fig{fig:Tree-TN-messages}c. Finally, note
that as $f_a(\Bx_a,\Bx'_a)$ are positive semi-definite, then
Eqs.~(\ref{eq:BP-i-to-alpha-DEFG}, \ref{eq:BP-alpha-to-i-DEFG})
imply that if the messages at time $t$ are positive semi-definite
then so are the messages at $t+1$.

The above discussion shows that like in the ordinary graphical
models, also here fixed points of the BP iterations are solving a
Bethe-Peierls type of approximation. In particular, defining the local
``marginals''
\begin{align*}
  P_a(\Bz_a) &\EqDef \sum_{\Bz\setminus \Bz_a} P(\Bz), &
  P_i(\Bz_i) &\EqDef \sum_{\Bz\setminus \{z_i\}} P(\Bz) .
\end{align*}
We can write a complex Bethe free-energy
\begin{align}
\label{eq:F-CBethe}
  F_{Bethe} = \sum_a \sum_{\Bz_a}P_a(\Bz_a)
    \ln\frac{P_a(\Bz_a)}{f_a(\Bz_a)}
     - \sum_i(d_i-1)\sum_{z_i}P_i(z_i)\ln P_i(z_i) , \qquad
     z_i=(x_i, x_i') ,
\end{align}
which is
defined by first choosing a specific branch of the logarithmic
function. Note that in this case, $d_i=|N_i|=2$ because there are
always exactly two adjacent factors to each variable $z_i$, and so
\begin{align}
  F_{Bethe} = \sum_a \sum_{\Bz_a}P_a(\Bz_a)
    \ln\frac{P_a(\Bz_a)}{f_a(\Bz_a)}
     - \sum_i\sum_{z_i}P_i(z_i)\ln P_i(z_i) .
\end{align}
It is not very hard to show that the even though $P_a(\Bz),
P_i(z_i), f_a(\Bz_a)$ might take complex values, $F_{Bethe}$ must be
real. In \Ref{ref:Cao2017-DEFG} it was argued that also in this case
fixed points of the BP iterations correspond to extremum points of
the above functional. We note, however, that unlike the ordinary
case, we see no reason why the complex Bethe free-energy should be
positive. Interestingly, in all of our numerics, it was positive.

\section{Proofs of Lemmas~\ref{lem:Lemma-I},\ref{lem:Lemma-II}}
\label{sec:proofs}

\subsection{Proof of Lemma~\ref{lem:Lemma-I}}

\begin{figure}
  \begin{center}
    \includegraphics[scale=1]{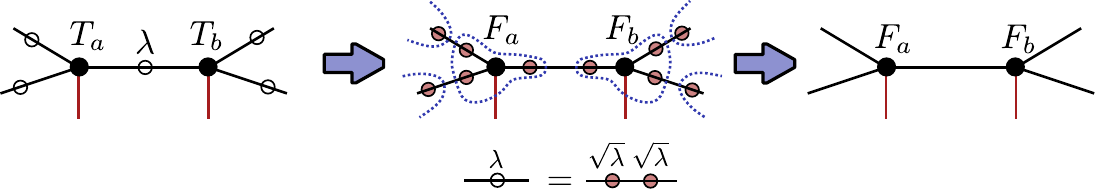}
  \end{center}
  \caption{ Swallowing the $\lambda$ weights in the $T$ tensors and
    obtaining an equivalent TN with $F$ tensors. The empty circles
    denote a Simple-Update weight tensor $\lambda_{x}\delta_{xy}$
    and the red circle denote its square root:
    $\sqrt{\lambda_{x}}\delta_{xy}$.} 
    \label{fig:swallow-lam}
\end{figure}

Assume a trivial-SU step changes the TN $\mcT$ to $\mcT'$ by locally
changing the adjacent tensors $T_a, \lambda, T_b$ to $T'_a,
\lambda', T'_b$, while keeping the rest of the tensors fixed (see
\Fig{fig:SU-step}a-f with trivial $U_{ab}=\Id$). To simplify the
book-keeping, we ``swallow'' the $\lambda$ tensors in the $T_a$
tensors by splitting every $\lambda$ tensor into $\lambda =
\sqrt{\lambda}\cdot\sqrt{\lambda}$ and contracting each
$\sqrt{\lambda}$ with its adjacent $T_a$ tensor, see
\Fig{fig:swallow-lam}. We denote the resulting tensor networks by
$\mcF, \mcF'$, and note that their local tensors are identical
except for the $F_a, F_b$ and $F_a', F'_b$ which are equal to the
$T_a, T_b, T_a', T_b'$ tensors contracted with the appropriate
$\sqrt{\lambda}$ tensors.  The fact that the contraction of $(T_a,
\lambda, T_b)$ is equal to the contraction of $(T_a', \lambda',
T_b')$ implies that the contraction of $(F_a, F_b)$ is equal to the
contraction of $(F_a', F_b')$. 
    
Let $\{m_{a\to b}(x,x')\}$ be fixed-point BP messages of $\mcF$. We
will use these messages to construct fixed-point BP messages
$\{m'_{a\to b}(x,x')\}$ of $\mcF'$ that give the same RDMs. All
messages except for the $a\to b$ and $b\to a$ messages remain the
same. The $a\to b$ and $b\to a$ messages are defined by the BP
iterative equations using the new tensors $F'_a, F'_b$ so that they
will satisfy them. For example, if tensor $F_a$ is connected also to
tensors $F_c,F_d$ in addition to $F_b$, then $m'_{a\to b}(x,x')$ is
given by the diagram in \Fig{fig:Tree-TN-messages}c, replacing $T$
tensors by corresponding $F'$ tensors. To finish the proof, we need
to show that this new set of messages (i) is a BP fixed-point, and
(ii) produces the same RDMs according to the BP formula (see
\Fig{fig:BP-RDM}). Clearly, for adjacent vertices that have
nothing to do with $a,b$, both conditions hold trivially, as the
relevant messages and underlying tensors are unchanged. Let us then
verify these points for vertices in the vicinity of $a,b$.
  \begin{figure}
    \begin{center}
      \includegraphics[scale=0.75]{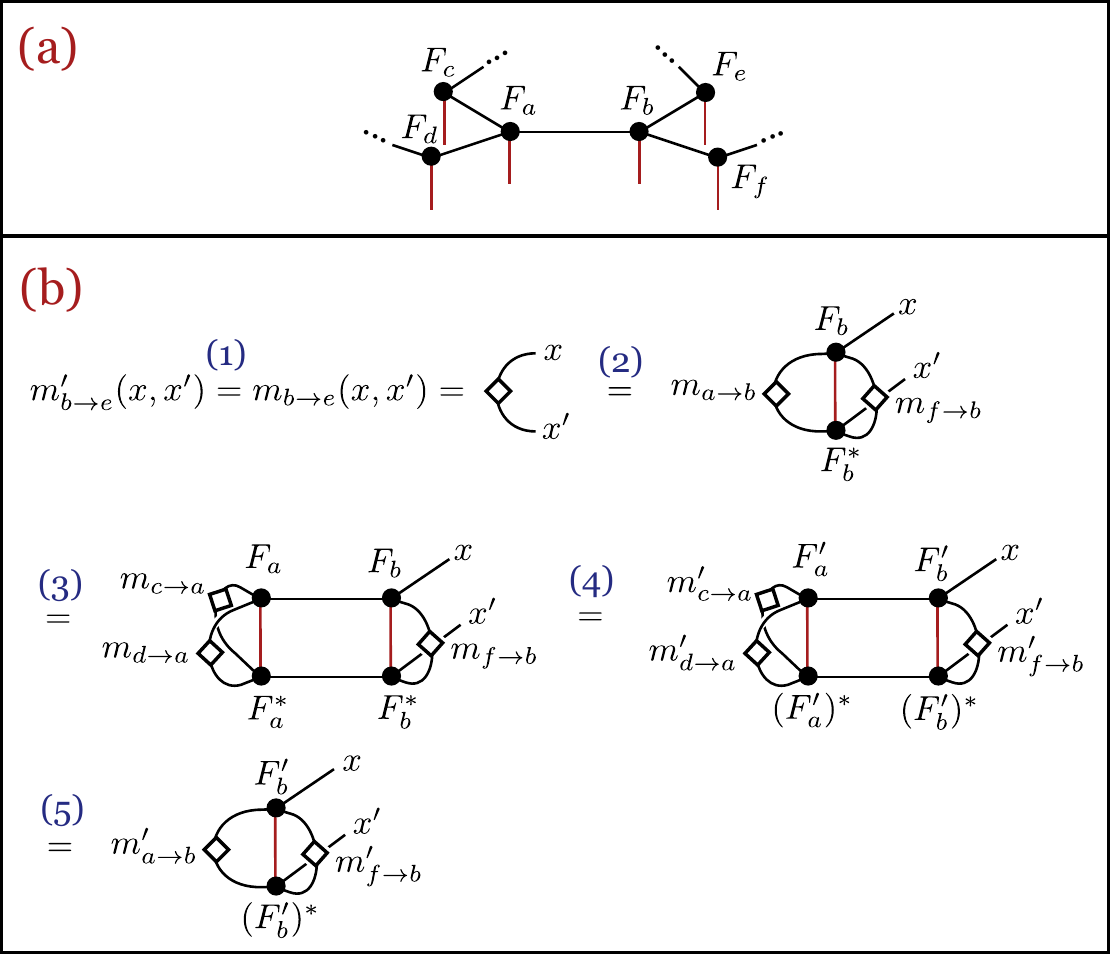}
    \end{center}
    \caption{    
      (a) The description of the TN.  canonical condition.
      (b) Proving that $m'_{b\to e}$ is given by the BP propagation
      of messages $m'_{a\to b}$ and $m'_{f\to b}$: Equality (1)
      follows from definition, $m'_{b\to e} = m_{b\to e}$. Then in
      (2) we use the assumption that $m_{b\to e}$ is a fixed point
      of the BP equation, and similarly in (3) we use that
      assumption on $m_{a\to b}$. In (4) we use the fact that the
      contraction of $F_a, F_b$ is equal to the contraction of
      $F_a', F_b'$, together with the definitions that all the new
      messages are equal to the old messages, except for the
      $a\leftrightarrow b$ messages. Finally, in (5) we use the
      definition of $m'_{a\to b}$ which was designed to satisfy the
      BP equations.}
      \label{fig:Lemma-I}
  \end{figure}

\paragraph{Checking point (i):} By definition, the $a\to b$ and
$b\to a$ messages satisfy the BP equations.  So we only need to
verify that other messages from $a$ or $b$ (but not between them)
satisfy the BP equations. Consider, for example, the message $b\to
e$ in \Fig{fig:Lemma-I}a.  We need to verify that $m'_{b\to
e}(x,x')$ is indeed a BP fixed-point, given as the appropriate
expression of $m'_{a\to b}, m'_{f\to b}$ (see
\Fig{fig:Tree-TN-messages}c for the BP equations).  This is proved
in \Fig{fig:Lemma-I}b in a series of 5 simple equalities (see the
caption for full explanation), which rely on the fact that the
original messages are fixed point of the BP equations, and that the
contraction of $F_a, F_b$ is equal to the contraction of $F'_a,
F'_b$.

\paragraph{Checking point (ii):} By definition if we are interested
in 2-body RDMs on vertices that are different from both $a$ and $b$,
then the RDM estimate will remain the same because neither the
relevant messages, nor the tensors changed. We only need to verify
for the RDM $\rho_{ab}$ and RDMs that contain $a$ or $b$ with other
adjacent node, such as $\rho_{be}$. For the former, $\rho_{ab} =
\rho'_{ab}$ because it depends on the incoming messages to the $a,b$
nodes (which remain the same), together with the contraction of
$F'_a, F'_b$, which by assumption is identical to that of $F_a,
F_b$. For the latter, the proof uses the same idea as in point (i).
Using the assumption that the contraction of $F_a, F_b$ is identical
to that of $F_a' F_b'$, and that $F_e'=F_e$, it is easy to show that
the 3-body RDM $\rho_{abe}$ is identical to that of $\rho_{abe}'$,
from which we deduce that $\rho_{be} = \rho'_{be}$.  This concludes
the proof of \Lem{lem:Lemma-I}.

\subsection{Proof of Lemma~\ref{lem:Lemma-II}}

\begin{figure}
  \begin{center}
    \includegraphics[scale=1]{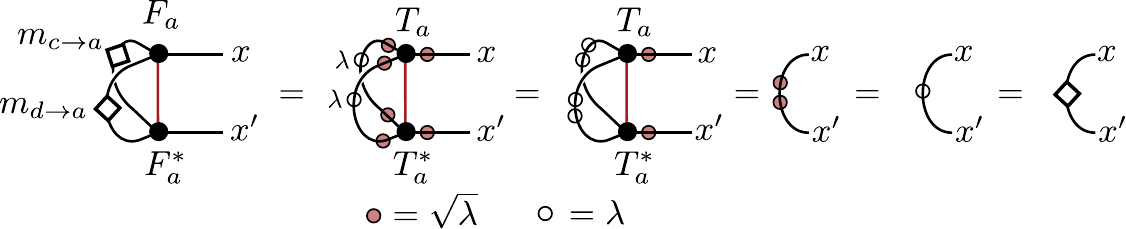}
  \end{center}
  \caption{The proof of \Lem{lem:Lemma-II}: defining the BP
  messages by \Eq{eq:canonical-BP} and using the canonical
  condition shows that these messages are fixed point of the BP
  equations.
    }
    \label{fig:Lemma-II}
\end{figure}

As in the first lemma, we first define $\mcF$ to be an equivalent TN
in which every $\lambda$ weight tensor in $\mcT$ was split into
$\sqrt{\lambda}\cdot\sqrt{\lambda}$ and the $\sqrt{\lambda}$ tensors
are contracted into the $T_a$ tensors to give the $F_a$ tensors (see
\Fig{fig:swallow-lam}). Next we define a set of messages 
  \begin{align}
  \label{eq:canonical-BP}
    m_{a \to b}(x,x') \EqDef \lambda_x \delta_{x,x'} 
  \end{align}
for every two adjacent vertices $a,b$, where $\lambda$ is the weight
on the $ab$ edge in the original $\mcT$ tensor. We claim that (i)
these messages are BP fixed-point on the $\mcF$ TN, and (ii) they
give the same 2-body RDMs as those of the trivial SU method of
quasi-canonical $\mcT$. Both claims are immediate. Claim (i) follows
by writing the BP equation for the $a\to b$ message in terms of the
$F_a$ tensor, and noticing that this expression is equal to it gives
the $\lambda_x \delta_{x,x'}$ using canonical condition on $\mcT$.
This is illustrated in \Fig{fig:Lemma-II}. Claim~(ii) follows from
definitions of the 2-body RDMs of the BP method and the SU method
(see \Fig{fig:SU-on-tree}e, and \Fig{fig:BP-RDM}).
  
\section{Numerical results}
\label{sec:extra-numerics}

\begin{figure}
  \begin{center}
    \includegraphics[scale=0.4]{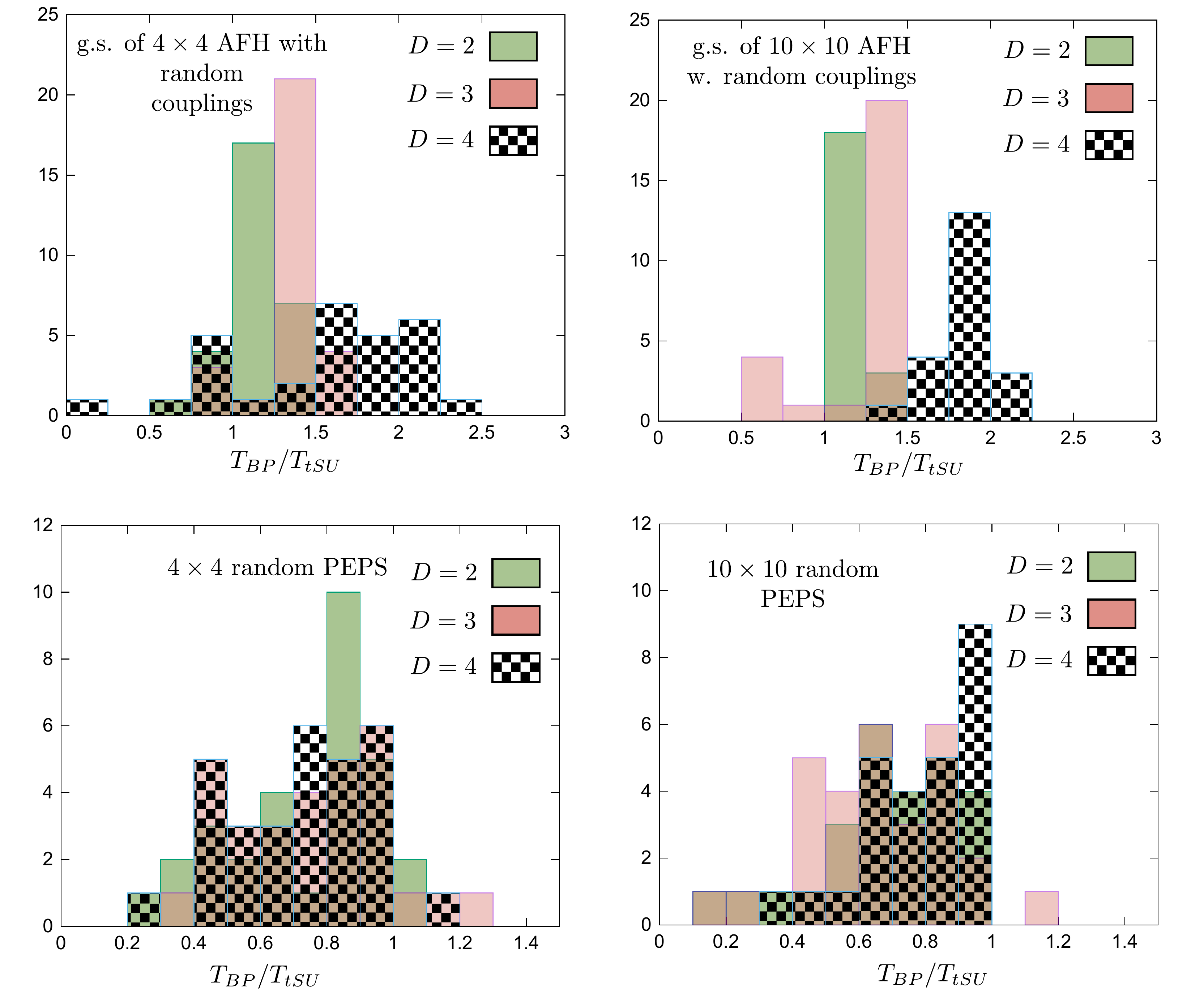}
  \end{center}
  \caption{The ratio of the number of iterations that takes the BP
  algorithm  to converge by that of the trivial-SU
  algorithm. The tests were on $4$ different systems: a $4\times 4$
  and $10\times 10$ anti-ferromagnetic Heisenberg models with random
  coupling, as well as $4\times 4$ and $10\times 10$ random PEPS.
  For one of these 4 cases, PEPS were used with bond dimension
  $D=2,3,4$, and the statistics was generated using $20-50$ different
  realizations. More details on the numerical procedure can be found
  the text body.}
  \label{fig:statistics}
\end{figure}

As part of this work, we ran simulations to compare the
convergence times of BP to the ones of trivial-SU on different PEPS
states. We tested the algorithms over two types of systems: $i)$
random PEPS, $ii)$ PEPS ground-states of the anti-ferromagnetic
Heisenberg model (AFH) with random couplings
\begin{align}
  H&=\sum_{\av{a,b}}J_{ab}\bm{\sigma}_a\otimes\bm{\sigma}_b, & 
  J_{ab}<0 .
\end{align}
Both systems were simulated on a $4\times 4$ and $10\times 10$
squared lattices.  In all tests, the physical bond dimension was
$d=2$ and virtual bond dimensions were $D=2, 3, 4$. All in all we
therefore tested $2\times 2\times 3 = 12$ different configurations.
For every configuration we used statistics of $20-50$ different
random realizations on which we did the analysis. 

In the random PEPS configurations, the tensor entries where chosen
as $a+ib$, where $a,b$ were uniformly distributed in $(-1, 1)$. In
the AFH configurations, we used random couplings $J_{ab}$ uniformly
distributed in the interval $(-1, 0)$. To obtain the ground states
of these models, we ran an imaginary time evolution with
simple-update, with decreasing values of imaginary time steps by
$\delta\tau=0.1,\dots , 0.0001$. After obtaining an approximation to
the ground state, we applied a random local gauge change on every
bond in order to get a TN that is far away from a canonical form.
Specifically, for every virtual edge $(a,b)$, we drew a random
matrix $V_{ab}$ which was a product of a random unitary with a
diagonal with random entries between $(0.5,2)$. We then inserted the
identity $V_{ab}^{-1}V_{ab}=\Id$ in the middle of the edge,
absorbing $V^{-1}_{ab}$ in $T_a$ and $V_{ab}$ in $T_b$. This way,
the resultant TN was far from quasi-canonical, yet represented the
same approximate ground state.

The convergence criteria for both BP and
trivial-SU was taken with respect to the averaged trace distance of
$2$-body RDMs between consecutive iterations (see
\Fig{fig:SU-on-tree}c for trivial-SU and \Fig{fig:BP-RDM} for BP
RDMs illustrations) such that $\frac{1}{m}\sum_{\av{a, b}}\norm{
\rho_{ab}^{(t+1)} -\rho_{ab}^{(t)}}_1 <10^{-6}$, where $\av{a,b}$
denotes nearest-neighbors nodes and $m$ is the total number of such
neighbors.

\end{document}